\documentclass[a4paper,twocolumn,showpacs,superscriptaddress,prl]{revtex4}
\usepackage[english]{babel}
\usepackage{graphicx}
\usepackage{xspace}
\usepackage{amsmath,amssymb}

\begin{document}

\title{Epitaxy of Fe$_3$O$_4$ on Si(001) by pulsed laser deposition
 using a TiN/MgO buffer layer}

\author{D.~Reisinger}
\email{Daniel.Reisinger@wmi.badw.de}
\affiliation{Walther-Mei{\ss}ner-Institut, Bayerische Akademie der
Wissenschaften, Walther-Mei{\ss}ner Str.~8, 85748 Garching,
Germany}

\author{M.~Schonecke}
\affiliation{Walther-Mei{\ss}ner-Institut, Bayerische Akademie der
Wissenschaften, Walther-Mei{\ss}ner Str.~8, 85748 Garching,
Germany}

\author{T.~Brenninger}
\affiliation{Walther-Mei{\ss}ner-Institut, Bayerische Akademie der
Wissenschaften, Walther-Mei{\ss}ner Str.~8, 85748 Garching,
Germany}

\author{M.~Opel}
\affiliation{Walther-Mei{\ss}ner-Institut, Bayerische Akademie der
Wissenschaften, Walther-Mei{\ss}ner Str.~8, 85748 Garching, Germany}

\author{A.~Erb}
\affiliation{Walther-Mei{\ss}ner-Institut, Bayerische Akademie der
Wissenschaften, Walther-Mei{\ss}ner Str.~8, 85748 Garching,
Germany}

\author{L.~Alff}
\email{Lambert.Alff@wmi.badw.de}
\affiliation{Walther-Mei{\ss}ner-Institut, Bayerische Akademie der
Wissenschaften, Walther-Mei{\ss}ner Str.~8, 85748 Garching, Germany}

\author{R.~Gross}
\affiliation{Walther-Mei{\ss}ner-Institut, Bayerische Akademie der
Wissenschaften, Walther-Mei{\ss}ner Str.~8, 85748 Garching,
Germany}

\date{received December 24, 2002}



\begin{abstract}
Epitaxy of oxide materials on silicon (Si) substrates is of great interest
for future functional devices using the large variety of physical
properties of the oxides as ferroelectricity, ferromagnetism, or
superconductivity. Recently, materials with high spin polarization of the
charge carriers have become interesting for semiconductor-oxide hybrid
devices in spin electronics. Here, we report on pulsed laser deposition of
magnetite (Fe$_3$O$_4$) on Si(001) substrates cleaned by an {\em in situ}
laser beam high temperature treatment. After depositing a double buffer
layer of titanium nitride (TiN) and magnesium oxide (MgO), a high quality
epitaxial magnetite layer can be grown as verified by RHEED intensity
oscillations and high resolution x-ray diffraction.
\end{abstract}
\maketitle


\section{Introduction}

Oxide thin films offer a large variety of physical properties
useful for future electronic devices. Much attention has been paid
to the ferroelectric, dielectric, and optical properties of
perovskite materials with respect to non-volatile memory
transistor applications or nonlinear optical devices \cite{Yu:00}.
In the area of so-called spintronics, ferromagnetic oxides with
high spin polarisation $P$ of the charge carriers at room
temperature are desired for spin injection semiconductor devices
as for example described by Datta and Das \cite{Datta:90}. While
it has been shown that the spin lifetime in semiconductors is high
as compared to metals \cite{Kikkawa:99}, one of the largest
problems is to create a spin polarized population of the charge
carriers in the semiconductor by injection of spin polarized
currents. The use of diluted magnetic semiconductors - if existent
- is possible but limited due to the low Curie temperatures of the
magnetic semiconductors so far \cite{Fiederling:99,Ohno:99}.
Therefore, half-metallic materials with a full spin polarization
of the charge carriers at the Fermi level and Curie temperatures
well above room temperature are under consideration for spintronic
devices. There are several candidates for materials with large $P$
close to 100\% such as the Mn-based Heusler alloys, the oxide
ferromagnets such as Fe$_3$O$_4$ \cite{Gupta:99} or CrO$_2$, the
doped manganites, and also the double perovskites
\cite{Philipp:01}. We note that the use of these materials for
ferromagnet-semiconductor hybrid devices requires the epitaxial
growth of these materials on semiconductors. This is a great
challenge with respect to materials technology. We further note
that the resistivity mismatch between ferromagnetic metals and the
semiconductors puts fundamental limits on the degree of spin
polarization achievable with spin injection \cite{Schmidt:00}.
The efficiency of spin injection, however, can be largely
increased by introducing a tunnel barrier between the
ferromagnetic metal and the semiconductor
\cite{Rashba:00,Fert:01}.

Here, we report on the fabrication and structural characterization of
magnetite (Fe$_3$O$_4$) epitaxial thin films on Si(001) substrates using a
double buffer layer of TiN and MgO. MgO is also a candidate for an
insulating barrier required for future tunneling injection devices. An
important step is the in-situ substrate treatment by an infrared laser beam
heating system. This allows for the controlled removal of the amorphous Si
oxide surface layer prior to film deposition monitored by reflection high
energy electron diffraction (RHEED).  After substrate cleaning, the thin
films are grown by ultra high vacuum (UHV) laser molecular beam epitaxy
(L-MBE) \cite{Gross:2000a,Klein:2001a}. Also, the complete growth process
is monitored by RHEED. We discuss in detail the growth of the buffer layer
system consisting of TiN and MgO matching the magnetite lattice constants,
and the epitaxial growth of Fe$_3$O$_4$ on top of the buffer layer system.
The surface of the completed Fe$_3$O$_4$ thin films as well as the
developing heterostructures is characterized after each process step by
{\em in situ} atomic force microscopy (AFM). The structural properties of
the completed hybrid heterostructures have been investigated by
high-resolution x-ray diffraction (HRXRD).

\section{Experimental Techniques}

\begin{figure}[tbh]
\begin{center}
 \includegraphics [width=0.9\columnwidth,clip]{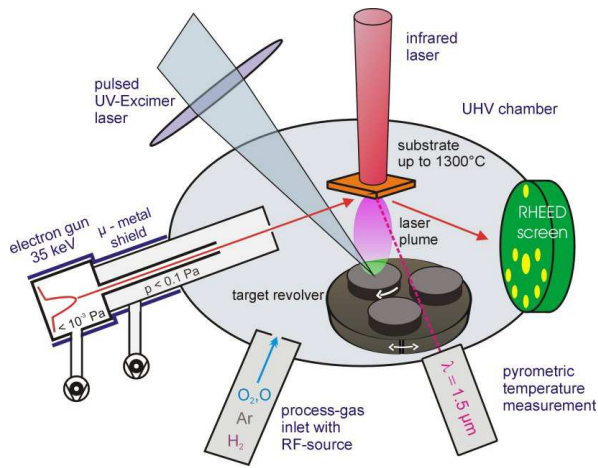}
 \caption{Sketch of the pulsed laser deposition chamber including the infrared laser heating system.}
 \label{fig1}
\end{center}
\end{figure}

The heteroepitaxial thin film structures have been grown using laser
molecular beam epitaxy (L-MBE) \cite{Gross:2000a,Klein:2001a}. The thin film UHV
system consists of three separate sub-chambers for pulsed laser deposition,
metallization, surface analysis and a load-lock. A sketch of the sub-chamber
for the pulsed laser deposition (PLD) is shown in Fig.~\ref{fig1}.
A technical description of the systems has been given
recently \cite{Gross:2000a,Klein:2001a}. Therefore, we only
discuss those parts of the system relevant for the work reported here. For
the present study, the two most important parts are (i) the
(high-pressure) RHEED system and (ii) the infrared laser heating system
used for the substrate heating.

\subsection{High pressure RHEED system}

Many oxide materials have to be grown in a large oxygen partial pressure
perturbing the RHEED analysis during growth due to the strong scattering of
the high energy electrons from the oxygen molecules. Therefore, a special
high pressure RHEED system has to be used, where the electrons travel only
a short distance in the high pressure deposition atmosphere, and the rest of
the beam path is pumped differentially \cite{Rijnders:97}. In our system a
double differential pumping unit is used allowing for the operation of the
RHEED system up to pressures of several 10\,Pa (for details see
Ref.~\cite{Klein:2001a}).

\subsection{Laser heating of Si(001) substrates}

An infrared laser heating system to heat substrates in a thin film
deposition system to temperature above $1200^\circ$C has first been
introduced by Ohashi {\em et al.} \cite{Ohashi:99}. Such a system has
several advantages with respect to thin film epitaxy. First, inside the
vacuum chamber of the deposition system the substrate is heated directly by
the infrared laser positioned outside the vacuum system. Therefore, the
substrate heating is more efficient than in the case of a radiation heater
positioned inside the vacuum chamber. In particular, the use of a radiation
heater results in parts with a temperature well above the substrate
temperature located inside the vacuum system. This in turn usually results
in reduced vacuum conditions. In our case, using the infrared laser
heating a substrate temperature $T_S$ well above 1200$^\circ$C can be
reached at a vacuum in the 10$^{-8}$\,mbar range. Compared to the
radiation heater setup in the same chamber, an improvement of the background pressure of about
an order of magnitude is achieved. Second, the laser heating system allows
for rapid changes of the substrate temperature, since only the substrate
is heated, which has a much smaller heat capacity than the system
radiation heater plus substrate. We realized changes of up to 500\,K/min,
only limited by the mechanical stability of the used substrate.
Third, high temperatures up to the melting point of Si at 1410$^\circ$C
can be reached without using an extensive amount of radiation shielding
preventing the optical access to the sample surface.

In our setup the beam of a diode laser with a wavelength of
950\,nm and a maximum power of 100\,W is fed through an optical
viewport and directly irradiates the backside of the Si(001)
substrate. Using transparent substrates like SrTiO$_3$, one has to
take into account the temperature dependence of the absorption
coefficient $\eta$. Here, a small amount of silver paint on the
backside of the substrate is sufficient to absorb most of the
laser power. This allows for the increase of the substrate
temperature to values, where $\eta$ increases and thus enables
substrate temperatures up to 1200$^\circ$C. In contrast, Si has
high infrared absorption already at room temperature and therefore
does not require any backside treatment. For temperature
measurement a high resolution pyrometer working at the wavelength
of $1.5\,\mu$m is used. The measured spot size is only 1~mm$^{2}$,
so even temperature inhomogeneities on the small substrates can
be detected. Changes in the emission coefficient $\varepsilon$ of the
substrates at the measuring wavelength give an error in true
temperature. Calibration measurements in an oven showed that this
error with a big change in $\varepsilon$ of 0.25 (worst case
scenario) is only about $\pm 3$\,K at $400^\circ$C and $\pm 20$\,K
at $1000^\circ$C. We note, however, that the for the deposition
processes important reproducibility of the deposition temperature
is with 5\,K on the used substrates much better. Compared to a
thermocouple the absolute exactness of this pyrometric measurement
is good and the reproducibility strongly improved.

\begin{figure}[tbh!]
\centering{%
\includegraphics [width=0.55\columnwidth,clip]{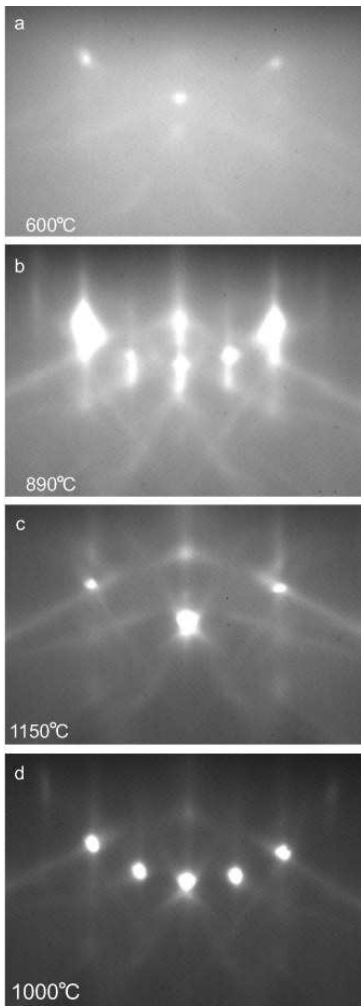}}
 \vspace*{0mm}\\
 \caption{RHEED images taken at [110] azimuth of the Si(001) substrate during the heat
treatment. (a) Untreated substrate at $T_S=600^{\circ}$C, (b) after heating
to $T_S=890^{\circ}$C, (c) at the maximum temperature of 1150$^{\circ}$C,
and  (d) after cooling to $T_S=1000^{\circ}$C.}
 \label{fig2}
\end{figure}

In the following we discuss the cleaning process for the surface
of the Si(001) substrates. In our work we used commercial Si(001)
single crystal wafers, which are one side polished and cut into
5\,mm by 5\,mm pieces. These pieces have been precleaned in
aceton and isopropanol using an ultrasonic bath. However, it is
unavoidable that a thin amorphous silicon oxide layer forms on the
substrate surface. For the epitaxial growth of a thin film
structure on the Si substrate, this amorphous oxide layer has to
be removed. Here, an {\em in situ} cleaning process is of great
advantage. For this purpose, the Si substrates have been laser
heated in vacuum to $T_S=900^{\circ}$C with a rate of 100\,K/min
and subsequently up to 1150$^{\circ}$C at 50\,K/min. Then, the
substrate was held for 10\,min at this temperature. After that is was cooled down
with a rate of 200\,K/min to the deposition temperature of
the first buffer layer (deposition of TiN at $T_S=600^{\circ}$).

The whole high temperature treatment of the substrate surface was monitored
by RHEED. The corresponding RHEED patterns are shown in Fig.~\ref{fig2}.
At low temperatures (Fig.~\ref{fig2}a, $T_S=600^\circ$), a large
homogeneous background signal is visible due to diffuse scattering from
the amorphous silicon oxide surface layer. Increasing the temperature, the
signal from the (2$\times$1) surface reconstruction becomes visible within
a lightly streaky pattern at about $T_S=900^\circ$ (Fig.~\ref{fig2}b). The
surface reconstruction takes place only after oxygen and other impurities
such as SiC have desorbed from the surface. It is interesting to note, that
at temperatures above 1100$^\circ$C the (2$\times$1) surface reconstruction
disappears. A possible explanation is that the dimer bonds of the (2$\times$1)
surface displacement structure are dissolved at that temperature. Note that
the diffuse background signal has disappeared, showing that the oxide has
been completely removed from the substrate surface. After cooling down
to $1000^\circ$C, a stable and clean (2$\times$1) surface is obtained with
strong diffraction spots on the Laue circle of 0$^{\rm th}$ order, faint
streaks, and Kikuchi lines. In summary, the process described above yields
a clean Si(001) (2$\times$1) surface, as indicated by the presence of the
additional diffraction spots, without any chemical surface treatment.

\section{Experimental Results and Discussion}
\subsection{Epitaxial growth of TiN, MgO, and Fe$_3$O$_4$}

For the pulsed laser deposition of the TiN, MgO, and Fe$_3$O$_4$ thin films
we used a 248\,nm KrF excimer laser with an energy density at the target of
2-5~J/cm$^{2}$ per shot, and a pulse repetition rate of 2-10\,Hz, depending
on the deposited material. Stoichiometric targets have been used for the
growth of all films.

\begin{figure}[tbh]
\includegraphics [width=1.0\columnwidth,clip]{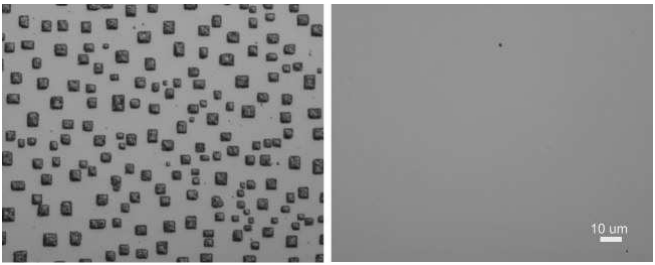}
 \vspace*{0mm}\\
 \caption{Optical micrographs of the surface of TiN films grown on Si(001).
Left: Island growth mode at $T_S=650^{\circ}$C. Right: Two-dimensional
growth mode at $T_S=600^{\circ}$C.}
 \label{fig3}
\end{figure}

We first discuss the TiN films deposited in an Ar atmosphere of
$2.5\cdot10^{-3}$\,mbar and a pulse repetition rate of 5\,Hz. As a
function of substrate temperature we have found two different
growth modes of the TiN(001) films on the Si(001) surface that
have been cleaned prior to deposition as described above. As
demonstrated by the optical micrographs shown in Fig.~\ref{fig3},
above about $T_S=650^\circ$C a three-dimensional Volmer-Weber
(island) growth mode is obtained, whereas below this temperature a
two-dimensional layer-by-layer growth mode is observed. For the
Volmer-Weber growth mode the islands are oriented along the [110]
direction of the Si(001) substrate. The size of the mostly
quadratic islands was about $3\,\mu\text{m}\times 3\,\mu\text{m}$.
It is not obvious why in the temperature range above $650^\circ$C
TiN grows with this preferential direction. We believe, that this
is due to the 45$^\circ$ rotation of the dimer rows in the
$2\times1$ surface reconstruction. With the lattice constant of Si
(diamond structure, $a=0.543$\,nm) the distance of the Si atoms
along the [110] direction is $a/\sqrt{2}=0.384$\,nm resulting in a
considerable lattice mismatch of 9.4\% to TiN (cubic,
$a=0.424$\,nm). The transition into the island growth mode with
increasing temperature itself is kinetic energy driven. For lower
temperatures, a smooth two-dimensional growth mode is obtained. It
is most likely that the growth occurs in the so-called
5-on-4-cube-on-cube bulk superstructure with a mismatch of about
-2.4\% \cite{Chowdury:94,Willmott:98}. In this superstructure 5
TiN unit cells overlap 4 Si(100) unit cells.

\begin{figure}[tbh]
\centering{%
\includegraphics [width=0.75\columnwidth,clip]{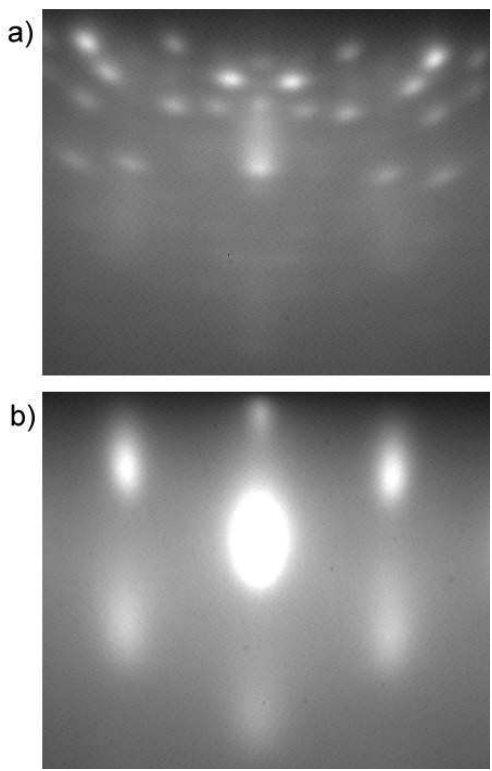}}
 \vspace*{0mm}\\
 \caption{RHEED pattern recorded during the growth of a TiN film on a Si(001)
substrate after 70 (a) and 1000 laser pulses (b). The deposition
temperature was $T_S=600^\circ$C.}
 \label{fig4}
\end{figure}

The RHEED pattern recorded during the growth of the TiN films is shown in
Fig.~\ref{fig4}. During the first 70 to 140 laser pulses corresponding to a film
thickness of about 0.4\,nm to 0.8\,nm, TiN grows in a smooth superstructure resulting
in a rich diffraction pattern (see Fig.~\ref{fig4}a). With increasing film thickness the fine
structure in the RHEED pattern disappears and clear diffraction spots are
observed at positions expected for small island transmission (see Fig.~\ref{fig4}b). This is consistent with
an {\em in situ} AFM analysis, which yields an average surface roughness below
1\,nm on a 1x1$\mu$m area at a film thickness of a few 10\,nm.

After the growth of the TiN buffer layer the substrate temperature is
lowered to $T_S=330^{\circ}$C. Then, as a second buffer layer MgO is
deposited at the same Ar pressure and a laser pulse repetition rate of
5\,Hz. During the MgO growth the RHEED pattern becomes more streaky. This
indicates a smoothening of the surface, which is consistent with {\em in
situ} AFM measurements of the MgO surface. On a $1\times 1\,\mu$m$^{2}$
area of the MgO surface the average roughness was about 0.6\,nm.
This shows that MgO is well suited as a buffer layer in the described
system since it provides smooth interfaces, and therefore can also serve
as insulator in a tunneling injection device.

\begin{figure}[tbh]
\centering{%
\includegraphics [width=0.95\columnwidth,clip]{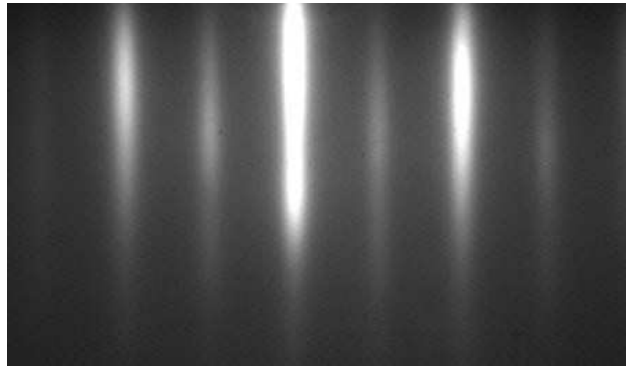}}
 \vspace*{0mm}\\
 \caption{RHEED pattern recorded during the growth of Fe$_3$O$_4$ on the
TiN/MgO buffer layer system after 5000 laser pulses corresponding to a
thickness of the  Fe$_3$O$_4$ film of about 40\,nm.}
 \label{fig5}
\end{figure}

On the completed TiN/MgO buffer layer system, an epitaxial magnetite film
has been grown at $T_S=330^{\circ}$C also in the same Ar atmosphere of
$2.5\cdot10^{-3}$\,mbar, at a laser pulse repetition rate of 2\,Hz.
These parameters have been found to result in the best RHEED intensity
oscillations due to a layer-by-layer growth of Fe$_3$O$_4$ on MgO
substrates \cite{Reisinger:03pre}. Note that the RHEED pattern recorded
during the growth of the magnetite film (see Fig.~\ref{fig5}) is streaky
but shows no indication of any transmission spots suggesting a smooth
surface in contrast to the TiN layer. The small roughness of the surface
of the magnetite film has been demonstrated directly by {\em in situ} AFM
measurements performed after deposition. For an area of
$1\times1\,\mu\text{m}^{2}$ a root mean square (rms) roughness of only
0.3\,nm has been found. The stripes in the RHEED pattern obtained for
magnetite have only about half the spacing of the diffraction spots
obtained for TiN and MgO. This is caused by the fact that the lattice
constant of magnetite is about twice the one of TiN and MgO (see
Figs.~\ref{fig4} and \ref{fig5}).

\begin{figure}[tbh]
\centering{%
\includegraphics [width=0.95\columnwidth,clip]{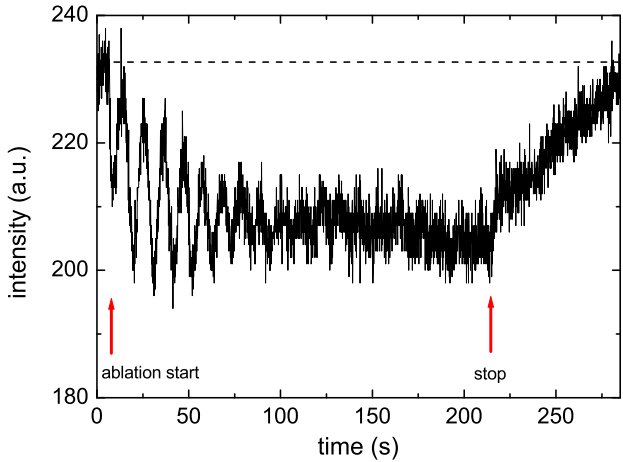}}
 \vspace*{0mm}\\
 \caption{RHEED oscillations of the (0,1) diffraction spot
recorded during the PLD growth of Fe$_3$O$_4$ on Si(001) at
$T_S=330^\circ$C. The decay of the amplitude of the RHEED oscillations due
to surface roughening and the recovery of the initial intensity after
stopping the deposition process is clearly visible. }
 \label{fig6}
\end{figure}

During the growth of magnetite on the TiN/MgO buffer layer system RHEED
oscillations could be observed as shown in Fig.~\ref{fig6}. From the
total number of RHEED oscillations, the number of laser pulses per single
oscillation and the film thickness determined from x-ray
reflectometry (see next section), we find that four RHEED oscillations are
obtained during the growth of a single unit cell thick layer. That is, the
magnetite unit cell grows in four sub-unit-cell blocks as discussed in
detail elsewhere \cite{Reisinger:03pre}.  After the growth of several unit
cells the amplitude of the RHEED oscillations decreases and finally
vanishes due to surface roughening. However, after an annealing step at
the deposition temperature, the RHEED intensity recovers to about the same
value observed before starting the deposition process. A key result is
that a layer-by-layer growth mode of Fe$_3$O$_4$ can be realized on the
TiN/MgO buffer layer system on Si(001) similar to the growth of
Fe$_3$O$_4$ on MgO single crystalline substrates.

\subsection{Structural analysis by HRXRD}

The structural properties of the thin films and the quality of epitaxy
have been characterized using a four circle x-ray diffractometer. We have
used an x-ray mirror and an asymmetric fourfold monochromator. This is
required to have sufficient intensity and resolution in order to be able to
distinguish between the diffraction peaks of the different layers which
are expected to be close together.

\begin{figure}[tbh]
\centering{%
\includegraphics [width=0.95\columnwidth,clip]{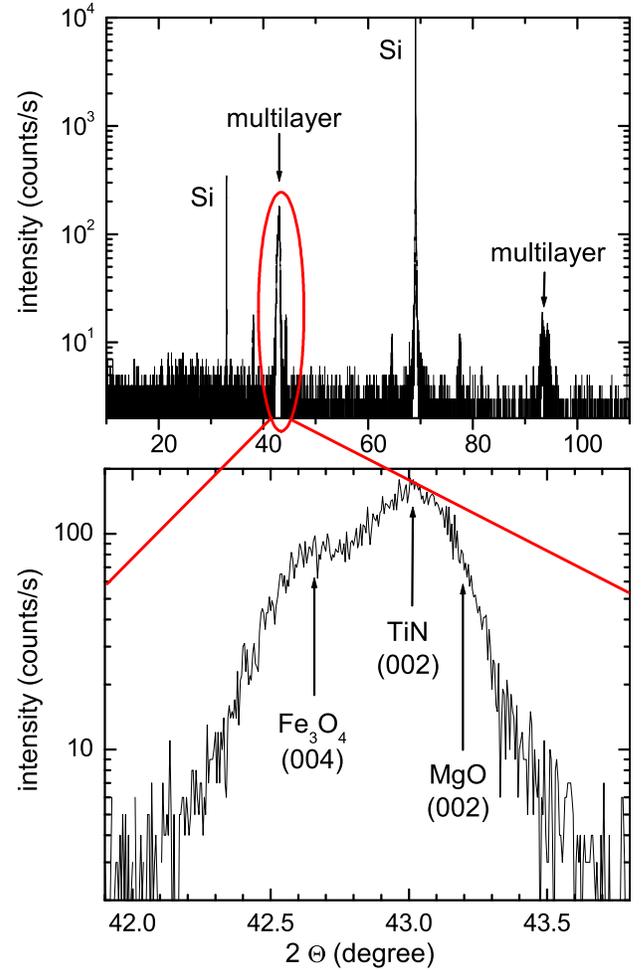}}
 \vspace*{0mm}\\
 \caption{2$\theta$-$\omega$ x-ray scans of the TiN/Mgo/Fe$_3$O$_4$
multilayer system grown on Si(001). The unlabeled peaks are also found in
the bare Si(001) substrate.  The lower panel shows part of the upper panel
on an enlarged scale from about 41.9$^{\circ}$ to 43.8$^{\circ}$.}
 \label{fig7}
\end{figure}

The upper panel of Fig.~\ref{fig7} shows a $\theta - 2\theta$ scan of the
completed TiN/MgO/Fe$_3$O$_4$ multilayer system grown on Si(001). All
layers have been grown in the ($00\ell$) orientation of the substrate
yielding only ($00\ell$) peaks. Within the resolution of our x-ray
diffraction system, no peaks of other chemical phases could be detected.
The detailed positions of the ($00\ell$) peaks of the thin TiN and MgO
buffer layers in the heterostructure are difficult to be determined (see
lower panel of Fig.~\ref{fig7}). Therefore, individual TiN films and TiN
films with thicker MgO top layers have been grown for x-ray analysis. From
these samples the following values for the $c$-axis parameter have been
found: Fe$_3$O$_4$: $c=0.847$\,nm, MgO: $c=0.419$\,nm, and TiN:
$c=0.420$\,nm. The FWHM of the rocking curve of the magnetite (004) peak
was about 0.7$^{\circ}$. This value is about ten times larger than the
value measured for Fe$_3$O$_4$ films grown directly on MgO substrates under
similar deposition conditions. This large difference can be attributed to
the large lattice mismatch between Fe$_3$O$_4$ (as well as the TiN/MgO
buffer layer) and Si compared to Fe$_3$O$_4$ and MgO (lattice mismatch of
Fe$_3$O$_4$: -2.4\% on Si(001) for the 5-on-4-cube-on-cube growth mode and
only -0.31\% on MgO). The large lattice mismatch results in strain
relaxation effects and the formation of misfit dislocations increasing
the mosaic spread of the films.

\begin{figure}[tbh]
\centering{%
\includegraphics [width=0.95\columnwidth]{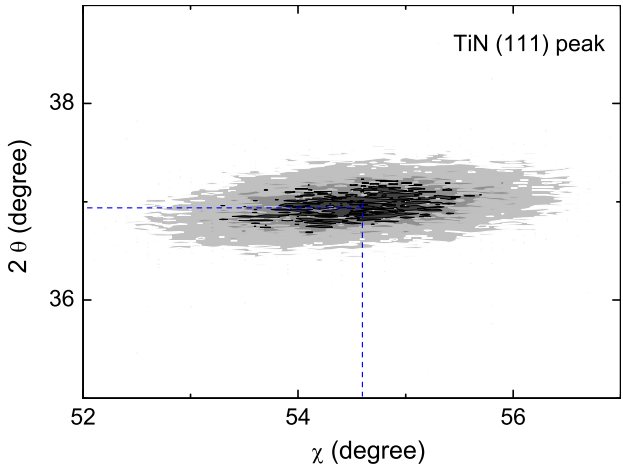}}
 \vspace*{0mm}\\
 \caption{X-ray $2\theta - \chi$ mapping of the TiN (111) peak.}
 \label{fig8}
\end{figure}

It is important to consider also the in-plane lattice constants in order
to estimate whether or not the strain due to the lattice mismatch between
the buffer layers and the substrate is (at least partially) relaxed. This
can be done by several methods. A 2$\theta - \chi$ mapping of the TiN (111)
peak is shown in Fig.~\ref{fig8}. Having determined the $c$-axis lattice
parameter ($c=0.420$\,nm) from the $\theta - 2\theta$ scans, the in-plane
lattice constant $a$ can be calculated for the tetragonal lattice of TiN
as

\begin{equation}
a = \sqrt{
\dfrac{h^{2}+k^{2}}{\left(\dfrac{2\sin\theta}{\lambda}\right)^{2}-\left(\dfrac{\ell}{c}\right)^{2}}
} \;.
 \label{xrayequ1}
\end{equation}

Here, $h$, $k$ and $\ell$ are the Miller indices of the respective
diffraction spot and $\lambda$ the wavelength of the Cu-K$_{\alpha 1}$
x-ray radiation used for the experiment. With $c= 0.420$\,nm, $h = k = \ell
= 1$, $\theta = 36.97^\circ$ and $\lambda = 0.15418$\,nm we obtain $a \simeq
0.422$\,nm. This value is close to the bulk value of TiN ($a=0.424$\,nm).
That is, most part of the TiN layer contributing to the intensity of the
(111) peak is relaxed. From the mapping shown in Fig.~\ref{fig8} the
in-plane lattice constant and the tetragonal distortion can be determined
both from the $\theta$-value using eq.(\ref{xrayequ1}) and the
$\chi$-value, which is given by the relation between the in-plane ($a=b$)
and out-of-plane ($c$) lattice constants as

\begin{equation}
\frac{c\sqrt{2}}{a}=\tan\chi \; .
 \label{xrayequ2}
\end{equation}

This relation directly follows from the symmetry of the tetragonal unit
cell. For a nondistorted lattice (cubic symmetry), the $\chi$-value should
be 54.74$^{\circ}$. As shown by Fig.~\ref{fig8}, the measured value is
54.64$^{\circ}$ resulting in a $c$- axis parameter of 0.42\,nm, which is
consistent with the value determined from the $\theta -2\theta $ scan. It
is evident from Fig.~\ref{fig8} that the (111) peak is much broader in the
$\chi$ than in $2\theta$ direction. This is caused by the measuring
geometry using a line profile of the x-ray beam. Analyzing the magnetite
(113) peak, we have determined the in-plane lattice constants of magnetite
to $a=b=0.832$\,nm by the same procedure. This value is close to the bulk
value of $a=0.835$\,nm.  We note that the relaxation of the TiN buffer layer
allows the growth of an almost strain free magnetite layer on Si(001).

We also performed $\varphi$ scans to check the epitaxial relations between
the different layers and the substrate. We found that the (111) TiN, (113)
magnetite, and (111) Si peak in a $\varphi$-scan (not shown) always appear
at the same in-plane angle $\varphi$ with the same fourfold symmetry. This
shows that the in-plane lattice vectors of all layers are strictly aligned
parallel to the Si substrate underlining the epitaxial nature of the
obtained growth. We obtain the following epitaxial relations: [100]$_{Fe_{3}O_{4}}$ $\parallel$
[100]$_{MgO}$ $\parallel$ [100]$_{TiN}$ $\parallel$ [100]$_{Si}$.

\begin{figure}[tbh]
\centering{%
\includegraphics [width=0.9\columnwidth]{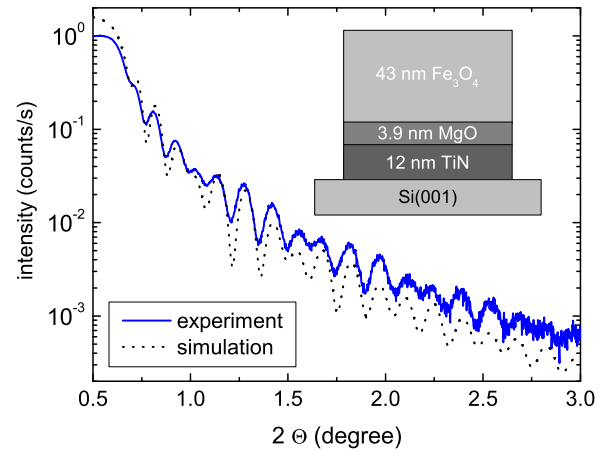}}
 \vspace*{0mm}\\
 \caption{X-ray reflectometry data obtained for a TiN/MgO/Fe$_3$O$_4$ multilayer
system grown on Si(001). Also shown is the simulation result.}
 \label{fig9}
\end{figure}

The thickness of the individual layers within the TiN/MgO/Fe$_3$O$_4$
multilayer structure was determined by low angle x-ray reflectometry. The
experimental results have been fitted using a simulation software based on
dynamical scattering theory \cite{refsim}. Fig.~\ref{fig9} shows a reflectometry
measurement of a complete multilayer system consisting of a
TiN/MgO/Fe$_3$O$_4$ multilayer on Si(001). Note the convincing agreement
between the simulation and the measurement of the complicated
heterostructure. From the refinement the layer thickness in this example
is obtained to 43\,nm for magnetite, 3.9\,nm for MgO, and 12\,nm for TiN.
By fitting the data also the roughness of the surface as well as of the
interfaces between the layers and the substrate could be estimated. The
derived roughness values are in the range bewteen 0.3 and 0.7\,nm. This is
fully consistent with the roughness values obtained from the AFM anaylsis.

\section{Summary}

Using pulsed laser deposition we have successfully grown magnetite
epitaxial thin films on Si(001) substrates using a double buffer layer
system consisting of TiN and MgO. For the reproducible cleaning of the Si
substrate we have used an {\em in situ} laser heating process with a
maximum temperature of 1150$^\circ$C. The substrate treatment and the
whole growth process was monitored directly by RHEED allowing the
determination of the surface structure at each process step. By x-ray
diffraction and AFM, the good crystalline quality of the multilayers and
the small roughness of the interfaces were verified. Due to their high
Curie temperature and predicted large spin polarization magnetite thin films on Si
substrates are highly interesting for potential applications in
spintronics, in particular for spin injection into semiconductors. The
compatibility of the presented multilayer structure with the insulating
material MgO may provide a way to fabricate a suitable tunnel barrier to
overcome the resistivity mismatch between ferromagnetic metals and
semiconductors.

This work was supported in part by the Deutsche Forschungsgemeinschaft
(project No.: Al/560) and the BMBF (project No.: 13N8279).

\end{document}